\documentclass[a4paper,10pt]{article}
\usepackage{stmaryrd}
\usepackage{amsfonts}
\usepackage{bbm}
\usepackage{amscd}
\usepackage{mathrsfs}
\usepackage{latexsym,amssymb,amsmath,amscd,amscd,amsthm,amsxtra,xypic}
\usepackage[dvips]{graphicx}
\usepackage[utf8]{inputenc}
\usepackage[T1]{fontenc}
\usepackage{lmodern}
\usepackage{amssymb}
\usepackage[all]{xy}
\usepackage{nicefrac,mathtools,enumitem}
\usepackage{microtype}
\usepackage{lipsum}
\usepackage{mathtools}

\usepackage{geometry}
\usepackage{hyperref}

\usepackage{times}
\usepackage{amsthm}
\usepackage{amsmath}
\usepackage{amsfonts}
\usepackage{amssymb}
\usepackage{mathdots}
\usepackage{color}
\usepackage{mathrsfs}
\usepackage{stmaryrd}
\SetSymbolFont{stmry}{bold}{U}{stmry}{m}{n}

\usepackage{bm}

\usepackage{tikz-cd}
\usepackage{tikz}
\usetikzlibrary{matrix,arrows,decorations.pathmorphing}
\usepackage[all]{xy}
\usepackage{graphicx}
\usepackage{subfigure}
\usepackage[toc,page]{appendix}
\usepackage[usestackEOL]{stackengine}

\newtheorem{thm}{Theorem}[section]
\newtheorem{lem}[thm]{Lemma}

\newtheorem{pro}[thm]{Proposition}
\newtheorem{ex}[thm]{Example}
\newtheorem{rmk}[thm]{Remark}
\newtheorem{defi}[thm]{Definition}

\newcommand{\be }{\begin{equation}}
\newcommand{\ee }{\end{equation}}

\newcommand{\pf}{\noindent{\bf Proof.}\ }
\newcommand {\sfPhi}{\mathsf{\Phi}}

\newcommand {\cl}[1]{\operatorname{cl}{(#1)}}

\newcommand {\onetwo}[1]{#1^{(1)}+#1^{(2)}}

\newcommand{\h}{\mathfrak h}

\newcommand {\mm}{\mathfrak m}

\def\qed{\hfill ~\vrule height6pt width6pt depth0pt}

\newcommand{\br}[1]{   [ \cdot,    \cdot  ]   }

\newcommand{\g}{\mathfrak g}

\newcommand{\Hom}{\mathrm{Hom}}

\newcommand{\Ker}{\mathrm{Ker}}

\newcommand {\IH}{{\rm Sect}}
\newcommand {\IR}{\mathbb{R}}
\newcommand {\C}{\mathcal C}

\newcommand {\GKZ}{_{\operatorname{\scriptstyle{gKZ}}}}
\newcommand {\GCKZ}{_{\operatorname{\scriptstyle{gcKZ}}}}

\newcommand {\UU}{\mathfrak U}

\newcommand {\eps}{\epsilon}
\newcommand {\scl}[1]{\operatorname{cl}{(#1)}}

\newcommand {\Ug}{U(\g)}
\newcommand {\fml}{[\negthinspace[\hbar]\negthinspace]}

\newcommand {\wh}{\widehat}
\newcommand {\IC}{\mathbb{C}}

\newcommand {\aand}{\qquad\text{and}\qquad}

\begin{document}
\title{\bf
Frobenius manifolds and quantum groups}
\author{
Xiaomeng Xu}

\newcommand{\Addresses}{{
  \bigskip
  \footnotesize
\noindent \textsc{Department of mathematics, ETH Zurich, 8092 Zurich, Switzerland
}\par\nopagebreak
  \textit{E-mail address}: \texttt{xiaomeng.xu@math.ethz.ch}
}}
\date{}
\maketitle 

\footnotetext{\it{Keyword}:  Semisimple Frobenius manifolds, Dubrovin connections, Knizhnik–Zamolodchikov equations, Isomonodromy deformation, Stokes matrices}


\maketitle

\begin{abstract}
We introduce isomonodromy Knizhnik–Zamolodchikov (KZ) connections with respect to the quantum Stokes matrices, and prove that the classical limit of the KZ type connections gives rise to the Dubrovin connections of semisimple Frobenius manifolds and flat F-manifolds. 
\end{abstract}
\setcounter{tocdepth}{1}
\tableofcontents

\section{Introduction}
The concept of Frobenius manifolds was introduced by Dubrovin \cite{Dubrovin} as a geometrical manifestation of the Witten-Dijkgraaf-Verlinde-Verlinde (WDVV) equations \cite{DVV, Witten0} governing deformations
of 2D topological field theories. Examples include Saito's Frobenius structures on unfolding spaces of singularities \cite{Saito2}, quantum cohomology (see e.g. \cite{Dubrovin, Manin}), the Barannikov-Kontsevich construction from
Batalin-Vilkovisky algebras \cite{BK} and so on.
The theory of Frobenius manifolds was investigated by many authors, and has been one of the principle
tools in the study of Gromov-Witten theory, integrable hierarchies, mirror symmetry, quantum singularity theory. The reader may consult \cite{CG, DuZhang, FJR, Lee, MO} and the literature cited there for a broader account of the subject. In particular, the works of Dubrovin \cite{Dubrovin, Dubrovin2}, Dubrovin-Zhang \cite{DuZhang}, Givental \cite{Givental, Givental2} and Boalch \cite{Boalch, Boalch3} connect the theory of semisimple Frobenius manifolds with meromorphic connections, integrable hierarchy, symplectic geometry on loop spaces and Poisson Lie groups, respectively. The main idea in the mentioned work of Dubrovin, Givental and Zhang is to construct all the building of a given 2D TFT starting from the corresponding Frobenius manifold, and two formalisms have been proposed: the Dubrovin-Zhang integrable hierarchy formalism and Givental’s quantization formalism. Both formalisms in turn rely on the theory of ODEs with irregular singularities.

A generalization of Frobenius manifolds, called F-manifolds, was introduced by Hertling and Manin \cite{HM}. Flat F-manifolds were studied by Getzler \cite{Get} and Manin \cite{Manin2}, and are related to the open WDVV equations \cite{HS} controlling the genus 0 open Gromov-Witten invariants, as remarked in \cite{BB}. Recently, the Givental’s reconstruction theory has been generalized to semisimple flat F-manifolds and F-Cohomological field theory by Arsie, Buryak, Lorenzoni and Rossi \cite{ABLR}.

Following \cite{Dubrovin}, semisimple Frobenius manifolds are characterized by certain flat connections, called the Dubrovin connections, and thus the space of Frobenius manifolds can be identified with the space $U_+$ of monodromy data of the Dubrovin connections. In a similar way, the space of flat F-manifolds are identified with the space $M_F$ of monodromy data of some other flat connections. See Section \ref{FFmanifold}. Ugaglia in \cite{Ugaglia} introduced a Poisson structure on $U_+$ by computing the pull-forward of the canonical linear Poisson structure under the 
Riemann-Hilbert maps (Dubrovin first computed the small rank case \cite{Dubrovin}). By generalizing the Aityah-Bott construction to the meromorphic connections, Boalch found an intrinsic approach to the Poisson structures on the general spaces of monodromy data, thus particularly on $U_+$ and $M_F$. Furthermore in \cite{Boalch3}, Boalch pointed out that $M_F$ is identified with the dual Poisson group $G^*$, while $U_+$ is the fixing locus of an algebraic Poisson involution on $G^*$, and the Poisson structure on $U_+$ is naturally induced by the Poisson involution. Then Xu \cite{XuP} pointed out that $U_+$ is naturally a Poisson homogeneous space. Here the (dual) Poisson groups are Lie groups with mutiplicative Poisson structures, introduced by Drinfeld \cite{Drinfeld2} as the classical limit of quantum groups. Similarly, the Poisson homogeneous space $U_+$ appears as classical limits of quantum symmetric pairs \cite{CGa}. 

Therefore the moduli spaces of semisimple Frobenius manifolds and flat F-manifolds have natural quantization from the perspective of quantum algebras and Poisson geometry (the irregular Atiyah-Bott construction \cite{Boalch3}). Thus there arise a natural question that if the quantum algebras correspond to certain deformations of Frobenius manifolds and flat F-manifolds (in a different category). In this paper and in \cite{Xu4}, we make the first step towards the question, by quantizing Dubrovin connections of Frobenius manifolds and flat F-manifolds via Knizhnik–Zamolodchikov type connections. We then explore the quantum analog of the monodromy data (particularly the Stokes matrices), isomonodromy deformation and Givental's symplectic actions from the theory of Frobenius and flat F-manifolds.

The quantization in this paper and \cite{Xu4} implies certain deformations of the datum of Frobenius manifolds, including the metric, Frobenius algebra structures and potential. At this moment, we only know that these (geometric) deformations are not in the category of Frobenius manifolds any more, and the defect are up to the "non-commutativity" in the quantum algebras. So an intrinsic geometric framework is remained to be understood. Besides, this paper doesn't include discussions about integrable hierarchies, an important part of the theory of Frobenius manifolds. The quantizaiton in this paper also implies certain deformations of the (dispersionless) principal hierarchy of a semisimple Frobenius manifold. We hope to explore these problems somewhere else.

\vspace{3mm}
The organization of the paper is as follows. The next section gives the preliminaries of Frobenius manifolds and flat F-manifolds, including some basic notions, the Dubrovin connections and moduli spaces. Section \ref{sec:isomono} introduces the isomonodromy (cyclotomic) KZ connections, and studies their canonical solutions and quantum Stokes data, and isomonodromy. Section \ref{sec:sclofKZ} proves that the classical limit of the isomonodromy (cyclotomic) KZ connection coincides with Dubrovin connections.

\section*{Acknowledgements}
\noindent
I give my warmest thanks to Anton Alekseev, Philip Boalch, Pavel Etingof, Huijun Fan, Victor Kac, Igor Krichever, Si Li, Nicolas Orantin and Valerio Toledano Laredo for the interest in our results and for useful
comments. This work was partially supported by the Swiss National Science Foundation grants P2GEP2-165118 and P300P2-174284. 

\section{Preliminaries on Frobenius manifolds and flat F-manifolds}\label{Frobeniusmanifold}
\subsection{Frobenius manifolds}
Let $M = (M, \mathcal{O}_M)$ be a complex manifold of dimension $n$. We denote by $\mathcal{T}_M$ its holomorphic tangent sheaf.
\begin{defi}\cite{Dubrovin}
A Frobenius manifold structure on $M$ is a tuple $(g,\circ, e, E)$, where $g$ is a non-degenerate $\mathcal{O}_M$-symmetric bilinear
form, called metric, $\circ$ is $\mathcal{O}_M$-bilinear product on $\mathcal{T}_M$, defining an associative and commutative algebra structure with the unit $e$, and $E$ is a holomorphic vector field on $M$, called
the Euler vector field, which satisfy:
\begin{itemize}
    \item $g(X\circ Y, Z)=g(X,Y\circ Z)$, $\forall X,Y,Z\in \mathcal{T}_M$;
    \item The Levi-Civita connection, denoted by $\hat{\nabla}$, with respect to $g$ is flat;
    \item The tensor $C:\mathcal{T}_M\rightarrow {\rm End}_{\mathcal{O}_M}(\mathcal{T}_M)$ defined by $C_XY=X\circ Y$ is flat;
    \item the unit element $e$ is flat.
    \item $\mathcal{L}_E g = D g$ for some constant $D$ and $\mathcal{L}_E (\circ) =\circ$.
\end{itemize}
\end{defi}

In particular, we have a structure
of Frobenius algebra $(\circ,g,e)$ on the tangent spaces $T_mM$ depending analytically on the point $m$.
This notion was introduced by Dubrovin as a geometrical/coordinate-free manifestation of the WDVV equations, see \cite{Dubrovin, Dubrovin2, Manin} for more details, examples and the relations with 2D topological field theories. It is also known as conformal Frobenius manifolds.

\subsection{Dubrovin connections}\label{Frobeniusmfld}
Given a Frobenius manifold $(M,\circ, g, e, E)$, let us take the sheaf ${\rm Pr}_M^*(TM)$ on $M\times (\mathbb{P}^1\setminus \{0,\infty\})$, where ${\rm Pr}:M\times (\mathbb{P}^1\setminus \{0,\infty\})\rightarrow M$ is the projection. The following
construction and proposition are known and can be found in different versions in \cite{Dubrovin, Manin}.
\begin{defi}\label{def:Duconn}
The Dubrovin connection $\nabla$ on ${\rm Pr}_M^*(TM)$ is defined for any local vector field $X\in \Gamma(TM)$ and $Y\in\Gamma({\rm Pr}_M^*(TM))$ by 
\begin{eqnarray}\label{structureconn}
\nabla_X Y &=& \hat{\nabla}_XY + zX\circ Y,\\
\nabla_{z}Y&=&\frac{dY}{dz}+\frac{1}{z}(\hat{\nabla}_YE-\frac{D}{2}Y)+E\circ Y.
\end{eqnarray}
\end{defi}
\begin{pro}
The connection $\nabla$ is flat.
\end{pro}
\pf It follows form the definition of Frobenius manifolds. For example, the flatness of the pencil of connections $\nabla_{X}(z)(Y) = \hat{\nabla}_{X}(Y) + zX\circ Y$ for any $X,Y\in\Gamma(TM)$  (viewed as parametrized by $z$) is equivalent to that the $(M,\circ,g)$ is associative and potential. \qed
\vspace{3mm}

The connection $\nabla$ is also known as the first structure connection in \cite{Manin}. 

\subsection{Semisimple Frobenius manifolds}\label{Dubrovinconn}
\begin{defi}\cite{Dubrovin}
A Frobenius manifold $(M,\circ, g, e, E)$ is called semisimple if for a generic point $m\in M$, the algebra $(T_mM,\circ,e)$ is semisimple, i.e., isomorphic, as a $\mathbb{C}$-algebra, to $\mathbb{C}^n$ with component-wise multiplication.
\end{defi}
The books \cite{Dubrovin, Manin} contain a complete review of these structures. Here we rewrite, without a proof, the Dubrovin connections of semisimple Frobenius manifolds in terms of the canonical coordinates.

\begin{pro}\cite{Dubrovin}
In a neighborhood $\C$ of a semisimple point $u_0$ of a Frobenius manifold $M$, there exist coordinates $u^1, ..., u^n$ such that 
\begin{itemize}
\item $\partial_i\circ\partial_j=\delta_{ij}\partial_i$, where $\partial_i:=\frac{\partial}{\partial u^i}$;

\item the eigenvalues of $E\circ$ at each point $m\in \mathcal{C}$ are $(u^1(m),...,u^n(m))$.
\end{itemize}
They are unique up to reordering and are called canonical coordinates. Furthermore,
\begin{itemize}
\item the metric $g$ is diagonal in the canonical coordinates, that is $g(u)=\sum_i h_i(u) du^i$,
for some nonzero functions $h_{1}(u), ..., h_{n}(u)$;
\item the unity vector field $e$ in the canonical coordinates has the form $e =\sum_i \partial_i$. 
\end{itemize}
\end{pro}

In the coordinates $(z,u^1,...,u^n)$ and in the frame of normalized idempotents $\{\frac{1}{\sqrt{h_i}}\frac{\partial }{\partial u^i}\}$, the Dubrovin connection of $M$ on $\IC\times\C$ can be written as follows: denote by the same letter $u={\rm diag}(u^1,...,u^n)$ the diagonal matrix and put the $n\times n$-matrix $V (u) := [(r_{ij}(u)),u],$ where $(r_{ij})$ is a $n\times n$-matrix with entries (the rotation coefficients for the canonical coordinates)
$r_{ij}:=\frac{\partial_j\sqrt{h_{i}}}{\sqrt{h_{j}}}, i\ne j,$ then
\begin{pro}\cite[Lemma 3.2]{Dubrovin2}\label{Dubrovineq}
The horizontal sections of the Dubrovin connection in Definition \ref{def:Duconn} satisfy
\begin{eqnarray}\label{classicaliso}
d_z F&=&\left(u+\frac{V(u)}{z}\right)Fdz,\label{Stokeseq}\\
\label{classicaliso1}d_\frak h F&=&\left( z d_\frak hU+\Lambda(u)\right) F,
\end{eqnarray}
where $d_\h$ is the de Rham differential on $\h$, and $\Lambda(u)$ is a matrix of 1-forms given by 
$\Lambda(u):=\sum_{i=1}^n V_idu^i$. Here $V_i:={\rm ad}_{E_{ii}}{\rm ad}^{-1}_U V(u)$ for $E_{ii}$ being the elementary matrix $(E_{ii})_{ab}=\delta_{ia}\delta_{ib}$. Furthermore, since the rotation coefficients are symmetric, i.e., $r_{ij}=r_{ji}$ (see \cite[Proposition 3.4]{Dubrovin}), then $V(u)$ is skew-symmetric.
\end{pro}

\begin{pro}\cite[Proposition 3.7]{Dubrovin}\label{V(u)property} For a semisimple Frobenius manifold, the function $V(u)$ in \eqref{Stokeseq} satisfies the system of equations
\begin{eqnarray}\label{simplepole}
\partial_kV (u)=[V(u), {\rm ad}_{E_{kk}}{\rm ad}^{-1}_U V(u)] , \ k = 1,...,n
\end{eqnarray}
where $E_{kk}$ is the elementary matrix $(E_{kk})_{ij} =\delta_{ik}\delta_{kj}$.
\end{pro}
One checks that equation \eqref{simplepole} ensures
the compatibility of the systems \eqref{Stokeseq} and \eqref{classicaliso1}, i.e., the flatness of the Dubrovin connection $\nabla$.

\subsection{Stokes phenomenon}\label{sec:sol}
\subsubsection{Canonical solutions and Stokes matrices}
For the moment, let us fix $u\in\h_{\rm reg}$ and concentrate on the equation \eqref{Stokeseq}. It has an irregular singularity (pole of order two) at $z=\infty$. The {\it Stokes rays} (also known as {\it anti-Stokes directions}) of the equation in the complex $z$-plane are the rays $\IR_{>0}
\cdot (u_i-u_j)^{-1}\subset\mathbb{C}$ for any $i\ne j$, determined by the canonical coordinates. Let us choose an initial Stokes ray $d_0$, and label the rays by $d_0,d_1,...,d_{2l}$ in a counterclockwise order. The {\it Stokes sector}
$Sect_i$ is the open regions of $\mathbb{C}$ bounded by the consecutive Stokes rays $d_i, d_{i+1}$. 

We choose the determination of ${\rm log}z$ with a cut along $d_0$. The following result is well–known. See, e.g.,
\cite[pp. 58-61]{Wasow} or \cite[Section 8]{Balser}.
\begin{thm}\label{jurk}
On each sector ${Sect}_i$, there
is a unique holomorphic function $H_i:{\rm Sect}_i\to {\rm GL}_n$ such that the function
\[F_i(z,u)=H_i z^{[V]} e^{zu}\]
satisfies equation \eqref{Stokeseq}, and $H_i$ can be analytically continued to the bigger sector $\wh{Sect}_i$ and then is asymptotic to $1$ within $\wh{\IH}_i$. Here $[V]$ denotes the diagonal part of $V$ (which is zero if $V$ is skew-symmetric), and $\wh{Sect}_i:=\left\{re^{ i \phi}|\,r\in {\rm Sect}_i, \phi\in(-\pi/2,\pi/2)\right\}.$
\end{thm}

We will call any solution $F_i$ (with the prescribed asymptotics in a Stokes sector) a canonical solution. In particular, $F_+:=F_0$ and $F_-:=F_l$ are canonical solutions (with respect to the initial sector ${\rm Sect}_0$ and its opposite sector $Sect_l$).
\begin{defi}
The {\it Stokes matrices} of the equation \eqref{Stokeseq} (with respect to
to the sector $Sect_0$) are the matrices $S_\pm(u,V)$ determined by
\[F_-(z,u)=F_+(z,u)\cdot e^{-2\pi i [V]} S_+ , \ \ \ \ \ 
F_+(z,u)=F_-(z,u)\cdot S_-e^{2\pi i [V]}
\]
where the first (resp. second) identity is understood to hold in $Sect_l$
(resp. $Sect_0$) after $ F_+$ (resp. $ F_{-}$)
has been analytically continued counterclockwise.
\end{defi}
\subsubsection{Extra symmetry}
Since the matrix $V(u)$ in \eqref{Stokeseq} is skew-symmetric, we have extra symmetry on the canonical solutions and Stokes matrices.
\begin{pro}\label{symmetry}\hfill

$(1)$ The canonical solutions $F_\pm(z,u)$ satisfy $F_\pm^T(-z,u)F_\pm(z,u) = 1.$ Here $T$ denotes the matrix transposition. 

$(2)$ The Stokes matrices satisfy $S_-=S_+^{-T}$.
\end{pro}
\pf See e.g. \cite[Proposition 3.10.]{Dubrovin} or \cite[Lemma 35]{Boalch}. \qed

\subsection{Isomonodromy deformations}
Following Jimbo-Miwa-Ueno \cite{JMU}, the system of PDEs \eqref{simplepole} describes the isomonodromy deformation of the meromorphic differential equation \eqref{Stokeseq}, which means that 

\begin{pro}\cite[Proposition 3.11]{Dubrovin}\label{sclisomono}
Along the solution leaf $V(u)$ of \eqref{simplepole}, the Stokes matrices $S_\pm(u)$ of \eqref{Stokeseq} are preserved (independent of $u$).
\end{pro}
As a consequence, the Stokes matrices of a semisimple Frobenius manifold is locally constant. It can also be seen by the following proposition.

\begin{pro}[\cite{Dubrovin}]\label{formalsol}
In a neighborhood $\C$ of a
semisimple point $u_0$, the canonical solutions $F_\pm(z,u)=H_\pm(z,u)e^{zU}$ of the first equation \eqref{Stokeseq} also satisfy the second equation \eqref{classicaliso1}. 
\end{pro}

\begin{rmk}
The equation \eqref{Stokeseq} has a regular singularity at $z=0$. In this paper, we are only concerned with the the Stokes/monodromy data at $z=\infty$. For the discussions about the solutions at $0$ and a complete set of the monodromy data (including also connections matrices, and monodromy around $z=0$), we refer the reader to \cite{Dubrovin}.
\end{rmk}

\subsection{Moduli space of semisimple Frobenius manifolds}\label{moduliofFM}

\subsubsection{Initial values of isomonodromy equations}\label{iviso}
Following the explicit constructure in \cite[Proposition 3.5]{Dubrovin} (for diagonalizable $V(u)$), any solution $V(u)$ of the system \eqref{simplepole} determines locally a semisimple Frobenius manifold. Together with Proposition \ref{V(u)property}, we have
\begin{thm}\cite[Corollary 3.3]{Dubrovin}
There exists a one-to-one correspondence between semisimple Frobenius manifolds (modulo certain transformations) and solutions of the system \eqref{simplepole}.
\end{thm}
The solution $V(u)$ of \eqref{simplepole} on a neighbourhood $\C$ of $u_0$, and therefore the corresponding semisimple Frobenius structure, is determined by the initial value at $u_0$: 
\begin{center}
{\em a skew-symmetric matrix $V_0(=V(u_0))$.}
\end{center}
Thus the matrices $V_0\in {\rm so}_n$ parametrize the semisimple germs at $u_0$, and we will denote by $\nabla_{V_0}$ the corresponding Dubrovin connection. 

\begin{rmk}
In other words, the germ of Frobenius
manifolds at a semisimple point $m\in M$ is determined by the linear data induced on the tangent space $T_mM$ and vice versa. Using the Dubrovin's flat connection formulation of Frobenius manifolds, it can also be seen from the deformation of connections studied by Malgrange \cite[Section 4.1]{Mal}, see also \cite{HM2}.
\end{rmk}

\subsubsection{Monodromy data}
One has no “natural” choice of a semisimple point $u_0$ in the Frobenius
manifold to specify the initial data.
However, due to the isomonodromy propertey (Proposition \ref{sclisomono}, see also \cite[Lecture 4]{Dubrovin2}), one can use the monodromy data, including the Stokes matrices, connection matrices and monodromy at $z=0$, to parametrize the Frobenius structure. The reconstruction of the Frobenius manifold from Dubrovin monodromy data is then given by solving certain Riemann-Hilbert-Birkhoff problem.

The most important part of the monodromy data is the Stokes matrices $S_+, S_-$ (generically they determine other parts of the monodromy data). Since we have the symmetry $S_-=S_+^{-T}$ (by Proposition \ref{symmetry}), the space of Stokes matrices from Frobenius manifolds is isomorphic to the space $U_+$ of upper triangular matrices with all diagonal entries being equal to $1$. In other words, the local moduli of semisimple Frobenius manifolds is identified with $U_+$, see \cite{Dubrovin} and \cite{CDG} for a detailed theory of the local invariants of the Frobenius structure.

We refer the reader to \cite{Manin} for the third description of moduli spaces: Manin's classification data via the second structure connections. 

\subsubsection{Riemann-Hilbert-Birkhoff correspondence and Dubrovin-Ugaglia Poisson spaces}
The above two descriptions of the local moduli of semisimple Frobenius manifolds are related by the Riemann-Hilbert-Birkhoff map $\mu$. At a chosen semisimple point $u_0$, the map is
\[\nu(u_0):{\rm so}_n^*\rightarrow U_+; \ V_0\mapsto S_+(u_0),\]
where $S_+(u_0)$ is the Stokes matrix of the equation \eqref{Stokeseq} at $u_0$, i.e., $\frac{dF}{dz}=\Big(u_0+\frac{V_0}{z}\Big)F$ with $V_0=V(u_0)$.

The induced Poisson structure on $U_+$, by the push forward of the canonical linear Poisson structure on ${\rm so}_n^*$, has been computed by Ugaglia \cite{Ugaglia} (and was computed earlier by Dubrovin for $n=3$), and is called Dubrovin-Ugaglia Poisson structure. Surprisingly, although the map $\nu(u_0)$ is highly transcendental, the push forward Poisson structure on $U_+$ is algebraic, and doesn't depend on the choice of $u_0$. 
The independence on $u_0$ has been interpreted by the Hamiltonian description of the isomonodromy deformation equation \eqref{simplepole}. The algebraic nature of the Poisson structure will be interpreted via quantum algebras in this paper.

Thus we have seen two parameterization of the semisimple germs, and the Riemann-Hilbert-Birkhoff map relate them and the corresponding Poisson structures.
\subsection{Flat F-manifolds}\label{FFmanifold}
\begin{defi}
A (homogeneous) flat F-manifold $(M, \nabla, \circ, e, E)$ is the datum of a complex manifold M, an
analytic connection $\nabla$ in the tangent bundle $TM$, an algebra structure $(T_mM, \circ)$ with unit $e$ on
each tangent space and an Euler vector field $E$, analytically depending on the point $m\in M$, such that the one-parameter
family of connections $\nabla_z=\nabla+z\circ$ is flat and torsionless for any $z\in\mathbb{C}$, and $\nabla e = 0$.
\end{defi}
Note that a flat F-manifold $M$ is a generalization of the notion of a Frobenius
manifold, where one loses the presence of a metric. However, similar to the case of Frobenius manifolds, these exist canonical coordinates near any semisimple point on $M$. In \cite{BB} the authors introduced
a Dubrovin type connection on $M\times \mathbb{C}^*$, and under the canonical coordinates the connection takes the same form as the Dubrovin connection in Proposition \ref{Dubrovineq}, except that $V(u)$ is not necessary skew-symmetric (or equivalently, one drops the symmetry of the rotation coefficients in the Darboux-Egorov system, see \cite{AL}). Furthermore, any flat F-manifold around a semisimple point can be obtained from such a connection. See e.g., \cite{AL} \cite[Section 1.4]{ABLR} for more details.

One can accordingly talk about the initial value of isomonodromy equations \eqref{simplepole} (where $V(u)$ is not necessary skew-symmetric), and introduce the Dubrovin monodromy data of flat F-manifolds, in a same way as in Section \ref{moduliofFM} for Frobenius manifolds. In particular, the germs of flat F-manifolds at a semisimple point $u_0$ can be identified with the space of Stokes matrices of 
the meromorphic linear system $\frac{dF}{dz}=\Big(u_0+\frac{V_0}{z}\Big)F$ for any $V_0\in {\rm gl}_n$, which is isomorphic to the Poisson Lie group ${\rm GL}_n^*$ dual to ${\rm GL}_n$. In an explicit term,
\[{\rm GL}_n^*=\{(b_-,b,t)\in B_-\times B\times \frak t~|~[b_-][b]=1, [b]=e^{\pi i t}\},\]
where $B$ and $B_-$ denote the upper and lower triangular Borel
subgroups of ${\rm GL}_n$, $\frak t$ is the set of $n$ by $n$ diagonal matrices, and $[b]$ takes the diagonal part of $b$. Note that the $\frak t$ component is determined by the $B$ or $B_-$ component, so for simplicity we will drop the $\frak t$ component. See e.g., \cite[Section 2]{Boalch} or \cite{LW} for the general theory of Poisson Lie groups.

Furthermore, the Riemann-Hilbert-Birkhoff map (at the semisimple point $u_0$) in this case becomes the Boalch's dual exponential map with a remarkable Poisson geometric property.

\begin{thm}\cite{Boalch}\label{BPoisson}
For the fixed $u$, the dual exponential map \[\nu(u_0):{\rm gl}_n^*\rightarrow {\rm GL}_n^* ; \ V_0\mapsto (S_-,S),\] associating the Stokes matrices $S_\pm(u_0,V_0)$ of 
the meromorphic linear system $
\frac{dF}{dz}=\Big(u_0+\frac{V_0}{z}\Big)F$ to any $V_0\in {\rm gl}_n\cong {\rm gl}_n^*$, is a local analytic Poisson isomorphism (provided ${\rm gl}_n^*$ is equipped with the canonical linear Poisson structure rescaled by $2\pi i$.)
\end{thm} 

As pointed out by Boalch \cite[Section 7]{Boalch}, the map $\nu(u):{\rm gl}_n^*\rightarrow {\rm GL}_n^*$ intertwines the two Poisson involutions $\tau:{\rm gl}_n\cong {\rm gl}_n^*; A\mapsto -A^T$, and $\tau':{\rm GL}_n^*\rightarrow {\rm GL}_n^*; (S_-,S)\mapsto (S^{T}, S_-^{T})$. The Poisson involution on ${\rm GL}_n^*$ naturally induces a Poisson structure on the fixing locus $U_+$, which coincides with the Ugaglia-Dubrovin Poisson structure. See \cite[Section 7]{Boalch} (or \cite{XuP} for the Poisson structures on the fixing locus induced by general Poisson involutions). It thus interprets the Dubrovin-Ugaglia Poisson structure by the standard Poisson structure on the dual Poisson Lie group, and when restricts to ${\rm so}_n\subset {\rm gl}_n$
\begin{thm}\cite{Ugaglia}\cite{Boalch}\label{DUPoisson}
The map $\nu(u):{\rm so}_n^*\rightarrow U_+$ is a Poisson map.
\end{thm}
\begin{rmk}
From a completely different viewpoint, and independently, Bondal discovered the same Poisson
structure on $U_+$ \cite{Bon}.
\end{rmk}

\section{KZ connections and isomonodromy deformation}\label{sec:isomono}
In this section, we study the Stokes phenomenon of (cyclotomic) KZ connections and their isomonodromy deformation. In Section \ref{sec:gKZ} and \ref{sec:scl}, we recall the notion of generalized KZ (gKZ) equations, and preliminaries on quantum algebras. In Sections \ref{sec:solgKZ}, \ref{sec:qStokes} and \ref{sec:isodeformation}, we study respectively the canonical solutions, quantum Stokes matrices and isomonodromy deformation of gKZ connections. Then in Section \ref{sec:StokesMa}, we introduce the notion of isomonodromy KZ connections, and in \ref{sec:StokesR} we show that the quantum Stokes matrices of KZ connections satisfy Yang-Baxter equations. In the last subsection \ref{sec:cycStokesMa}, we summarize the Stokes phenomenon of cyclotomic KZ connections studied in our previous paper \cite{Xu4}, which is parallel to the results in Sections \ref{sec:solgKZ}--\ref{sec:StokesR} for KZ connections.

\subsection{Generalized KZ equations}\label{sec:gKZ}
Let us take the Lie algebra $\g={\rm gl}_n$, and $\frak h\subset\g$ the set of diagonal matrices. Let $\Omega:=\sum e_\alpha\otimes e_a$ for any orthonormal basis $\{e_a\}$ of $\g$. Set $\UU=\Ug\fml$ and denote by $\wh{\otimes}$ the
completed tensor product of $\IC\fml$--modules.  

The {\it generalized KZ (gKZ) connection} \cite{FMTV} with a parameter $u\in\frak h$ is the $\UU^{\wh{\otimes} 2}$–valued connection on $\mathbb{C}$ given by
\begin{eqnarray}\label{dKZ}
\nabla\GKZ=d_{z}-\Big(u^{(2)}+\hbar\frac{\Omega}{z}\Big) dz.
\end{eqnarray}
We would like to introduce the solutions of $\nabla\GKZ F_\hbar=0$ with prescribed asymptotics at $z=\infty$, as well as certain $\hbar$-adic property. For that let us first recall the notion of admissibility. 

\subsection{Admissibility and classical limit}\label{sec:scl}
Let $\eps:\UU\to\IC\fml$ be the counit of $\UU$. Then $\UU$ splits as
$\Ker(\eps)\oplus\IC\fml\cdot 1$, with projection onto the first summand
denoted by $\pi$. 
Define $\UU' \subset\UU$ by
\[\UU' =\{x\in\UU|\, \pi^{\otimes n}\circ\Delta^{(n)}(x)\in\hbar^n\UU ^{\otimes n},\,n\geq 1\},\]
where $\Delta$ is the coproduct on $\UU$, and $\Delta^{(n)}:\UU \to\UU ^{\wh{\otimes} n}$
is the iterated coproduct recursively defined by
$\Delta^{(1)}={\rm id}$, and $\Delta^{(n)}=(\Delta\otimes{\rm id}^{\otimes (n-2)})\circ
\Delta^{(n-1)}$ for $n\geq 2$. 
The algebra $\UU $ has a natural Hopf algebra structure, known as a quantum formal series Hopf algebra, and has the following well--known facts (see e.g. \cite{Ga}).

\begin{lem}\label{hbaradic}
We have $\UU =U(\hbar\frak g\fml)$. That is, $x=\sum_{n\geq 0}\hbar^n x_n$ lies
in $\UU' $ if, and only if the filtration order of $x_n$ in $U(\frak g)$ is less than or
equal to $n$. Furthermore, $\UU' $ is a flat deformation of the completed symmetric algebra $\wh
{S}\frak g=\prod_{n\geq 0}S^n\frak g$.
\end{lem}
An element $A\in\UU\wh{\otimes}\UU$ is called admissible, if $A$ is further inside the subalgebra $\UU' \wh{\otimes}\UU$.
Given an admissible $A\in\UU'\wh{\otimes}\UU$, the {\it classical limit}
of $A$, denoted by $\cl{A}$ is the image of $A$ in
\[\UU' \wh{\otimes}\UU/\hbar(\UU' \wh{\otimes}\UU)
=\wh{S}\frak g\wh{\otimes}\Ug\]
Given that $\wh{S}\frak g=\IC\llbracket \frak g^*\rrbracket $, we will regard $\cl{A}$ as formal
function on $\frak g^*$ with values in $\Ug$. 

Since $\UU' $ is a flat deformation of $\wh{S}(\frak g)$, it follows that any $A\in \UU' \wh{\otimes}\UU$ give rise to an element $A^\circ$ of $(\wh{S}(\frak g) \wh{\otimes} U(\g))\fml$, called the renormalization of $A$. It can be computed as follows: by the Poincar${\rm \acute{e}}$-Birkhoff-Witt isomorphism $U(\g)\cong S(\frak g)$, $A$ is regarded as an element of $(S(\frak g)\otimes U(\g))\fml$, i.e., a function $A(x)$ on $x\in\frak g^*$
with values in $U(\g)\fml$, then 
\begin{eqnarray}\label{PBW}
A^\circ(x)=A(\hbar^{-1}x)=\cl{A}+O(\hbar).
\end{eqnarray}
Conversely, one get an admissible element in $\UU' \wh{\otimes}\UU$ from any element of $(\wh{S}(\frak g) \wh{\otimes}\UU)\fml$.

Now let us introduce group like elements in the classical limit. Let $G={\rm GL}_n$, regarded as an affine algebraic group over $\IC$, and denote by $\IC[G]$ its ring of regular functions. Let $\IC [\frak g^*]$ be the algebra of regular
functions on $\frak g^*$, and $\frak m\subset \IC [\frak g^*]$ the ideal of $0\in\frak g^*$. For any positive integer $k$, we denote by $G(\IC [\frak g^*]/\mm^p)={\rm Alg}_\IC(\IC[G], \IC [\frak g^*]/\mm^p)$ the set of $\IC [\frak g^*]/\mm^p$-points of $G$, and by $G(\IC [\frak g^*]/\mm^p)_\mm$ the normal subgroup
\[G(\IC [\frak g^*]/\mm^p)_\frak m = \{\phi\in{\rm Alg}_\IC(\IC[G], \IC [\frak g^*]/\mm^p) ~|~ \phi(I)\subset\frak m \}\] for $I = \{f\in \IC[G]~|~ f(1) = 0\}$ being the augmentation ideal. 
Let $\IC[[G]]=\lim\IC[G]/I^n$
be the completion of $\IC[G]$ at the identity, then $\Ug$ is identified, as
a Hopf algebra, with the continuous dual
$\IC[[G]]^*=\left\{\varphi\in\Hom_\IC(\IC[G],\IC)|\,\varphi(I^n)=0, n\gg 0\right\}$.
Thus $G(\IC [\frak g^*]/\mm^p)_\mm$ embeds into the Hopf
algebra $ (\IC [\frak g^*]/\mm^p)\otimes \Ug$ over $\IC [\frak g^*]/\mm^p$, and elements in its image are group like. 
Therefore,
the inverse limit \[G\llbracket\frak g^*\rrbracket_0:=\underleftarrow{\rm lim}~ G(\IC [\frak g^*]/\mm^p)_\mm\] embeds into the
topological Hopf algebra $\widehat{S}(\frak g)\wh{\otimes}\Ug=\lim_p(\IC [\frak g^*]/\mm^p)\otimes \Ug$.

\subsection{Canonical solutions of generalized KZ equations}\label{sec:solgKZ}
Following \eqref{PBW}, the renormalization of any admissible function $H_{\hbar}$ on a complex manifold $X$ leads to an expansion \[H^\circ_{\hbar}=H+\hbar H_{1}+\hbar^2 H_{2}+\cdots,\] 
where each $H_{i}\in \widehat{S}(\frak g) \wh{\otimes}\Ug$. 
We say that the function $H_\hbar:X\to \UU'\wh{\otimes}\UU$ is holomorphic if each term $H_i:X\to \widehat{S}(\frak g) \wh{\otimes}\Ug$ (thus the truncation by any finite order of $\hbar$) of $H^\circ_\hbar:X\to(\widehat{S}(\frak g) \wh{\otimes}\Ug)\fml$, as a formal function on $X\times \g^*$ valued in $U(\g)$, is actually a holomorphic function on $X\times \g^*$ valued in a finite dimensional space.

Let us denote the right and left half planes by ${\rm Sect}_+$ and ${\rm Sect}_-$ respectively, and denote the super sectors by $\wh{Sect}_\pm:=\left\{re^{ i \phi}|\,r\in {\rm Sect}_\pm, \phi\in(-\pi/2,\pi/2)\right\}$. As we will see, they are the Stokes sectors and supersectors of the gKZ equations with $u\in\h_{\rm reg}(\mathbb{R})$ (the set of diagonal matrices with purely imaginary eigenvalues). We fix henceforth the branch of $\log z$ with a cut along the ray $i\mathbb{R}_{<0}$.

\begin{thm}\label{th:Stokes}
For any $u\in\frak h_{\rm reg}(\mathbb{R})$, there is a unique holomorphic admissible function
$H_{\hbar\pm}:{\rm Sect}_\pm\to\UU'\wh{\otimes}\UU$ such that the function
\[{F_{\hbar\pm}}(z,u)=H_{\hbar\pm}(z,u)\cdot z^{\hbar\Omega_0}\cdot e^{zu^{(2)}}\]
satisfies $\nabla\GKZ{F_{\hbar\pm}}=0$, and $H_{\hbar\pm}(z,u)$ can be analytically continued to $\wh{Sect}_\pm$ and tends to $1$
as $z\to \infty$ within $\wh{Sect}_\pm$. Here $\Omega_0:=\sum_i h_i\otimes h_i$ for any orthonormal basis $\{h_i\}$ of $\frak h$.
\end{thm}
\pf First recall that the PBW isomorphism
between the
symmetric algebra $S(\g)$ and the universal enveloping algebra $U(\g)$, as a vector space, is given by the symmetrisation map $\sigma: S(\g)\rightarrow U(\g)$. Thus the multiplication on $U(\g)$ can be transported to $S(\g)$ as an associative star product
product
\begin{eqnarray}
x\star y =\sum_{p=0}^\infty B_p(x, y)\hbar^p,
\end{eqnarray}
where $B_p$ is the homogeneous component of degree $-p$ of the map 
\[B: S(\g)\otimes S(\g)\rightarrow S(\g); \ B(x, y) = \sigma^{-1}(\sigma(x)\sigma(y)),\] 
and for each $p\ge 0$, $B_p$ is a bidifferential operator of order $\le q$. For any formal function $F\in \widehat{S}(\frak g) \wh{\otimes}\Ug$ and $x\in\g$, we denote by $B_j(x,F)\in \widehat{S}(\frak g) \wh{\otimes}\Ug$ the resulting formal function (where $B_j(x,\cdot)$ acts on the first component of $F$).

Now let $H_{\hbar}(z)$ be an admissible function depending on the complex parameter $z$, and let $H^\circ_{\hbar}=H_0+\hbar H_{1}+\hbar^2 H_{2}+\cdots$.
We denote by $\Omega\star H_i$ (resp. $H_i\star \Omega_0$) the product of $\Omega\in S(\g)\otimes U(\g)$ (resp. $\Omega_0$) and $H_i$, given by the star product on the $S(\g)$ component and the product on $U(\g)$, and denote by $\Omega H_i$ the product given by the symmetric algebra product on the $S(\g)$ component. Then for any $i\ge 0$, we have
\begin{eqnarray}
\Omega\star H_i=\Omega H_i+\sum_{j\ge 1} H_{i,j}\hbar^j\in (\wh{S}(\frak g) \wh{\otimes} U(\g))\fml,
\\
H_i\star\Omega_0=H_i\Omega_0+\sum_{j\ge 1}{ _j}H_{i}\hbar^j\in (\wh{S}(\frak g) \wh{\otimes} U(\g))\fml,
\end{eqnarray}
where $H_{i,j}(u,z)=\sum_a e_a\cdot B_j(e_a,H_i)$ and $_jH_i(u,z)=\sum_iB_j(H_i,h_i)\cdot h_i\in \widehat{S}(\frak g) \wh{\otimes}\Ug$, for $\{e_a\}$ being an orthonormal basis of $\g$ and $\{h_i\}$ being an orthonormal basis of $\frak h$.

A direct computation shows that
\begin{lem}
The gKZ equation for $F_{\hbar}(u,z)=H_{\hbar}(z,u)\cdot z^{\hbar\Omega_0}\cdot e^{zu^{(2)}}$ is equivalent to the following equations of the functions $H_i(u,z)$ for any $i\ge 0$,
\begin{eqnarray}\label{recursiveeq}
\frac{dH_i}{dz}=u^{(2)}H_i+\frac{\Omega H_i}{z}-\Big(H_iu^{(2)}+\frac{H_i\Omega_0}{z}\Big)+\sum_{k=0}^{i-1} \frac{H_{k,i-k}}{z}-\sum_{k=0}^{i-1} \frac{_{i-k}H_{k}}{z}.
\end{eqnarray}
\end{lem}
Thus our goal it to find holomorphic solutions $H_i$ of equation \eqref{recursiveeq} with the asymptotics $H_0(z,u)\rightarrow 1$ and $H_i(z,u)\rightarrow 0$ as $z\rightarrow \infty$, within $\wh{{\rm Sect}}_+$. Here recall that holomorphic means that the function $H_i\in\widehat{S}(\frak g) \wh{\otimes}\Ug$ coincides with (the formal Taylor series with respect to $A\in\g^*$ at $A=0$ of) a homomorphic function on ${\rm Sect}_+\times \g^*$ valued in a finite dimensional space.

First let us solve the equation for $i=0$. For any $A\in\g^*\cong\g$, we denote by $A_0\in\h$ the diagonal part of $A$. Let $H_0(u;A):Sect_+\rightarrow G=GL_n$ be the unique holomorphic solution of 
\begin{eqnarray}
\frac{dH_0}{dz}=\Big(u+\frac{A}{z}\Big)H_0-H_0\Big(u+\frac{A_0}{z}\Big)
\end{eqnarray}
with the asymptotics $H_0(z,u;A)\sim 1$ as $z\rightarrow \infty$. Let us regard $H_0(z,u)$ as a holomorphic function of $A\in\g^*$ with $H_0(z,u;A=0)=1$, and denote by the same letter
\[H_0(u):Sect_+ \longrightarrow G\llbracket\frak g^*\rrbracket_0\]
its formal Taylor series with respect to $A\in\g^*$ at $A=0$. Then one checks that $H_0(u)\in G\llbracket\frak g^*\rrbracket_0\subset \widehat{S}(\frak g) \wh{\otimes}\Ug$ is the required holomorphic solution of the equation \eqref{recursiveeq} for $i=0$.

In the following, we solve recursively the equation \eqref{recursiveeq}. Let us denote by 
\[U_0(\g)\subset U_1(\g)\subset U_2(\g)\subset\cdot\cdot\cdot\] the the standard order filtration of $U(\g)$ given by $\deg(x)=1$ for $x\in\g$. First we assume that $H_i(u)$ are the unique holomorphic solutions of \eqref{recursiveeq} for $i=0,...,k-1$ with the prescribed asymptotics, and assume that $H_0^{-1}H_i$ is actually a holomorphic solution on ${\rm Sect}_+\times \g^*$ valued in $U_i(\g).$ Since \eqref{recursiveeq} is inhomogeneous, in the following we will prove that exists a function $T$, such that $H_k:=H_0 T$ is a solution of equation \eqref{recursiveeq} for $i=k$.

Since $B_j(e_a,\cdot):{S}(\frak g) \rightarrow {S}(\frak g)$ is a differential operator of order $\le j$ and $H_0\in G\llbracket\frak g^*\rrbracket_0$ is group valued, then $H_0^{-1} H_{0,k}$ and $H_0^{-1}\cdot _{k}H_0$ are holomorphic functions on ${\rm Sect}_+\times \g^*$ valued in $U_{k}(\g)$. Similarly, by the assumption on $H_0^{-1}H_i$ for any $0\le i\le k-1$, we see that $H_0^{-1}H_{i,k}$ and ${H_0^{-1}}\cdot \prescript{}{k}{H}_i$ are valued in $U_{k}(\g)$. Thus if we denote by $X_{k}(z)$ the function \[H_0^{-1}\Big(\sum_{j=0}^{k-1} \frac{H_{j,k-j}}{z}-\sum_{j=0}^{k-1} \frac{_{k-j}H_{j}}{z}\Big)\in \widehat{S}(\frak g) \wh{\otimes}\Ug\] 
on ${\rm Sect}_+$, then $X_i$ is a holomorphic function on $\wh{Sect}_+\times \g^*$ valued in $U_{k}(\g)$.

\begin{lem}\label{lem:inhom}
For any $A\in\g^*\cong\g$, there exits a unique holomorphic function $T(z;A):\wh{\rm Sect}_+\rightarrow U_k(\g)$ satisfying the equation (in $\wh{\rm Sect}_+$)
\begin{eqnarray}\label{inhomeq}
\frac{dT}{dz}+\Big[T,u+\frac{A_0}{z}\Big]=X_{k}(A),
\end{eqnarray}
and such that $T(z;A)\sim 0$ as $z\rightarrow\infty$.
\end{lem}
\pf Without loss of generality, let us assume $A_0=0$. Let $(\mathcal{B}X_k)(\omega;A)$ be the Borel transform of the function $X_k(z;A)$ of $z$, which is holomorphic in $\omega\in {\rm Sect}_+$ (see e.g., \cite[Section 5.2]{Balser}). Let $(\mathcal{B}T)(\omega;A)\in U_k(\g)$ be a function of one complex variable $\omega\in {\rm Sect}_+$ satisfying the system of linear equation
\begin{eqnarray}\label{Boreleq}
\mathcal{B}T\cdot u-(u+\frac{1}{\omega})\cdot \mathcal{B}T=\mathcal{B}X_k.
\end{eqnarray}
Here $\mathcal{B}T\cdot u$ is the product of $\mathcal{B}T\in U_k(\g)$ and $u\in\g\subset U(\g)$ in $U(\g)$, and note that the left hand side of \eqref{Boreleq} is still in $U_k(\g)$.
The coefficient matrix of the linear system is only singular at $\omega=\frac{1}{\lambda-\eta}$ for any two different eigenvalues $\lambda, \eta$ of the adjoint action $u\in\h_{\rm reg}(\mathbb{R})$ on $U_k(\g)$. Thus the function $\mathcal{B}T$ is well defined and holomorphic within ${\rm Sect}_+$. Besides, since the Laplace transform (see \cite[Section 5.1]{Balser}) of $\mathcal{B}X_k$ is the function $X_k$ holomorphic in $\wh{\rm Sect}_+$, the Laplace transform $\mathcal{L}(\mathcal{B}T)$ of $\mathcal{B}T$ is also holomorphic within $\wh{\rm Sect}_+$. Moreover one checks that $T(z;A):=\mathcal{L}(\mathcal{B}T)(z)$ is a solution of \eqref{inhomeq} in $\wh{\rm Sect}_+$.

We remark that the above function $(\mathcal{B}T)(\omega;A)$ is actually the Borel transform of a formal solution $\hat{T}(z;A)$ of \eqref{inhomeq}. In more explicit terms, if we take the asymptotic expansion $X_k(z;A)\sim \sum_{i\ge 1}\frac{f_i}{z^i}$ at $z=\infty$ within ${\rm Sect}_+$, and let $\hat{T}(z;A)=\sum_{i\ge 1}\frac{T_i}{z^i}$ be the formal solution of the formal equation \eqref{inhomeq} with the formal coefficients $\sum_{i\ge 1}\frac{f_i}{z^i}$ replacing $X_k(A)$ on the right hand side, then the series $(\mathcal{B}T)(\omega;A):=\sum_{i\ge 1}\frac{T_i}{(i-1)!\omega^i}$ satisfies the equation \eqref{Boreleq}, and (as long as $\omega$ keeps a positive distance from the two singular directions $i\mathbb{R}_{\lessgtr 0}$) the Laplace transform of $(\mathcal{B}T)(\omega;A)$ is a holomorphic solution \eqref{inhomeq} in a proper sector. \qed

\vspace{2mm}
Now we denote by the same letter $T(u):\wh{\rm Sect}_+ \longrightarrow \wh{S}(\g)\wh{\otimes}U_{k}(\g)$
the formal Taylor series of the function $T(z;A)$ in Lemma \ref{lem:inhom} with respect to $A\in\g^*$ at $A=0$. Then $T(u)$ satisfies
\[\frac{dT}{dz}+\Big[T,u^{(2)}+\frac{\Omega_0}{z}\Big] =X_{k},\]
and one checks that $H_{k}=H_0T$ is a solution of equation \eqref{recursiveeq} for $i=k$ with the required asymptotics. 

Thus we have constructed the solutions $H_i$ of \eqref{recursiveeq}. Let $H_{\hbar+}$ be the function whose renormalization is $H^\circ_{\hbar+}=H_0+\hbar H_1+\cdots$, then $H_{\hbar+}$ is the desired function in ${\rm Sect}_+$. In a same way, one can construct $H_{\hbar-}$ in ${\rm Sect}_-$. \qed

\begin{rmk}
Note that for general $u\in\h_{\rm reg}$ (not necessary real), the coefficient matrix of the linear system \eqref{Boreleq} in $U_k(\g)$ is singular at values of $\omega$ not necessary purely imaginary. Thus in general there will be infinite many singular directions for the resummation of the formal solution, as the integer $k$ goes to infinite. In this paper, we focus on the case $u\in\h_{\rm reg}(\mathbb{R})$ which simplifies the story, but it is interesting to study the general cases with possible infinite many singular directions/Stokes rays, and particularly the isomonodromy deformation in this setting.
\end{rmk}

\begin{rmk}\label{diffsolutions}
Solutions with prescribed asymptotics of gKZ equations were first given by Toledano Laredo in \cite{TL} in different approach and setting, where the notion of holomorphic function in the infinite dimensional space $\UU^{\wh{\otimes}2}$ is different from ours. It is interesting to compare these two constructions.
\end{rmk}
\subsection{Quantum Stokes matrices}\label{sec:qStokes}
Let $F_{\hbar\pm}$
be the canonical solutions of $\nabla\GKZ F_\hbar=0$ with respect to ${\rm Sect}_\pm$. 

\begin{defi}
For any $u\in\h_{\rm reg}(\mathbb{R})$, the quantum Stokes matrices $S_{\hbar\pm}(u)\in\UU^{\wh{\otimes} 2}$ of the gKZ connection \eqref{dKZ} are defined by
\[{F_\hbar}_+={F_\hbar}_-\cdot e^{-\hbar\pi i\Omega_0} S_{\hbar+}(u)\aand
{F_\hbar}_-={F_\hbar}_+\cdot S_{\hbar-}(u)e^{\hbar\pi i\Omega_0}\]
where the first identity is understood to hold in $\IH_-$ after ${F_\hbar}
_+(z,u)$ has been continued across the ray $\IR_{\geq 0}$, and the second
in $\IH_+$ after ${F_\hbar}_-(z,u)$ has been continued across $\IR_{\leq 0}$. 
\end{defi}
\begin{rmk}
In the categorical setting, the canonical solutions and Stokes matrices of gKZ equations were studied in \cite{Xu3,Xu4}. Besides, (confluent) hypergeometric type solutions of gKZ equations for finite dimensional representation spaces were given in \cite{FMTV}. These solutions have different asymptotics at $z=\infty$ in different sectors $\IH_\pm$ (and differ with the canonical solutions by a constant connection matrix). 
Then the Stokes matrices should be computed by comparing the different asymptotics. It is interesting to get an integral expression of the Stokes matrices (thus R-matrices in representation spaces, see \cite{Xu3, Xu4} or Section \ref{sec:StokesR}) along this line, and we expect that the computation is closely related to the theory of canonical bases. 
\end{rmk}
\subsection{Isomonodromy deformations}\label{sec:isodeformation}
In this subsection, we study the isomonodromy deformation problem of the gKZ equation, that is to find a function $\Omega(u)\in \UU^{\wh{\otimes} 2}$ on a neighbourhood $\mathcal{D}$ of $u_0$ in $\h_{\rm reg}(\mathbb{R})$ with $\Omega(u_0)=\Omega$, such that the Stokes matrices $S^{u_0}_\pm(u)$ of the connection \begin{eqnarray}\label{dKZ1}
\nabla^{u_0}=d_{z}-\Big(u^{(2)}+\hbar\frac{\Omega(u)}{z}\Big) dz.
\end{eqnarray} 
are preserved (independent of $u$). We will use the superscript to stress the dependence on the initial point $u_0$.

First we check the dependence of the canonical solutions on $u$. We take a root space
decomposition $\g = \mathfrak h \oplus_{\alpha\in\Phi}\mathbb{C}e_\alpha$, and for any positive root $\alpha\in\Phi_+$ set $C_{\alpha}=e_\alpha e_{-\alpha}+e_{-\alpha}e_\alpha\in \UU$. 
\begin{lem}\label{dependu}
The canonical solutions $F_{\hbar\pm}$ satisfy
\[\left(d_\h-\frac{\hbar}{2}\sum_{\alpha\in\Phi_+}\frac{d\alpha}{\alpha}
\Delta(C_\alpha)-zd_\h u^{(2)}\right){F_\hbar}_\pm=
{F_\hbar}_\pm\left(d_\h-\frac{\hbar}{2}\sum_{\alpha\in\Phi_+}\frac{d\alpha}{\alpha}
(\onetwo{C_\alpha})\right)\]
where $\Delta$ is the standard coproduct on $\UU$, and $C_\alpha^{(1)}:=C_\alpha\otimes 1\in\UU^{\wh{\otimes} 2}$, $C_\alpha^{(2)}:=1\otimes C_\alpha\in\UU^{\wh{\otimes} 2}$.
%
\end{lem}
The proposition motives the following notion of isomonodromy Casimir elements. Let us consider the equation
\begin{eqnarray}\label{gauge1}
d_\frak h T(u)=\frac{\hbar}{2}\sum_{\alpha\in\sfPhi_+}\frac{d\alpha}{\alpha}(\onetwo{C_\alpha})T(u).
\end{eqnarray} 
Let $T(u)\in \Ug^{\wh{\otimes} 2}$ be the holomorphic solution of \eqref{gauge1} defined on a neighbourhood $\C$ of $u_0$ in $\h_{\rm reg}(\mathbb{R})$ with initial condition $T(u_0)=1\otimes 1$ (see Section \ref{sec:hadic} for an expression of $T(u)$). Let $\Omega$ be the Casimir element (which has the form $\Omega=\sum e_\alpha\otimes e_a$ for any orthonormal basis $\{e_a\}$ of $\g$). 

\begin{defi}
The function $\Omega(u):\mathcal{D}\rightarrow\UU^{\otimes 2}~;~ u\mapsto T(u)^{-1}\Omega T(u)$ is called the isomonodromy Casimir (with respect to the chosen initial point $u_0$).
\end{defi}

Let us then consider the connection \eqref{dKZ1} with residue the isomonodromy Casimir $\Omega(u)$.
\begin{pro}\label{canonicalsol}
For any $u\in\mathcal{D}$, let $H_{\hbar\pm}:\IH_\pm\to \UU'{\wh{\otimes}}\UU$ be the holomorphic functions as in Theorem \ref{th:Stokes}.
Then $\mathcal{H}_{\hbar\pm}(z,u):=T(u)^{-1}H_{\hbar\pm}(z,u)T(u)$ are the unique holomorphic functions on $\IH_\pm$ valued in $\UU'{\wh{\otimes}}\UU$, such that $\mathcal{H}_{\hbar\pm}(z,u)$ tends to $1$
as $z\to \infty$ within $\wh{{\rm Sect}}_\pm$, and the
function
\[{\mathcal{F}_\hbar}_\pm(z,u)=\mathcal{H}_{\hbar\pm}(z,u)\cdot z^{\hbar\Omega_0}\cdot e^{zu^{(2)}}\]
satisfies the equation $\nabla^{u_0} \mathcal{F}_\hbar=0$.
\end{pro}
\pf Due to the fact $[T(u),z u^{(2)}]=0$ and $[T(u),\Omega_0]=0$, we have \[{\mathcal{F}_\hbar}_\pm(z,u)=T(u)^{-1}H_{\hbar\pm}(z,u)T(u) z^{\hbar\Omega_0} e^{z u^{(2)}}=T(u)^{-1} F_{\hbar\pm}(z,u) T(u).\] Here $F_{\hbar\pm}$ are the canonical solutions in Theorem \ref{th:Stokes}. It then follows from Theorem \ref{th:Stokes} and Equation \eqref{gauge1} that for any $u\in\mathcal{D}$, the functions ${\mathcal{F}_\hbar}_\pm(z,u)$ satisfy the equation $\nabla^{u_0} \mathcal{F}_\hbar=0$.

The fact that $\mathcal{H}_{\hbar\pm}(z,u)$ is valued in $\UU'\wh{\otimes}\UU$ is a consequence of the $\hbar$-adic property of $T(u)$ given below in Section \ref{sec:hadic}. Finally, the asymptotic behaviour and uniqueness of $\mathcal{H}_{\hbar\pm}$ follows from those of $H_{\hbar\pm}$.
\qed

\vspace{2mm}
Similar to Section \ref{sec:qStokes}, we can introduce the quantum Stokes matrices of $\nabla^{u_0}$, then we have
\begin{thm}\label{thm:qisomono}
The quantum Stokes matrices $S^{u_0}_{\hbar\pm}(u)\in \in\UU^{\wh{\otimes} 2}$ of the connection \eqref{dKZ1}, with residue the isomonodromy Casimir $\Omega(u)$, stay constant in $\mathcal{D}$ (independent of $u$). That is $S^{u_0}_{\hbar\pm}(u)=S_{\hbar\pm}(u_0)$.
\end{thm}
\pf It follows directly from the definition of Stokes matrices and Lemma \ref{dependu}.
\qed 
\subsection{Isomonodromy KZ connections}\label{sec:StokesMa}
Given any $u_0\in\h_{\rm reg}(\mathbb{R})$, denote by $\Omega(u)$ the associated isomonodromy Casimir on a neighbourhood $\mathcal{D}\subset \h_{\rm reg}(\mathbb{R})$, we introduce
\begin{defi}\label{def:IsoKZ}
The isomonodromic KZ (iKZ) connection, with respect to the chosen initial point $u_0$, is the $\UU^{\wh{\otimes}2}$–valued connection on $\times\IC\times\mathcal{C}$ given by
\begin{eqnarray}
&&\nabla^{u_0}= d_z-\left( u^{(2)}+\hbar\frac{\Omega(u)}{z}\right)dz,\label{isoKZ1}\\
&&\nabla_{d}^{u_0}=d_\frak h-\left( z d_\h u^{(2)}+\hbar\sum_{\alpha\in\sfPhi_+}\frac{d\alpha}{\alpha}T^{-1}(u)C_\alpha T(u)\right).\label{isoKZ2}
\end{eqnarray}
\end{defi}
Here the subscript $d$ in $\nabla_{d}^{u_0}$ stands for "dynamical connection", a notion borrowed from \cite{FMTV}. One checks that the iKZ connection is flat: the defining equation \eqref{gauge1} for $T(u)$ is nothing other than the integrability condition for the iKZ connection. 

\begin{pro}\label{solISO}
The functions $\mathcal{F}_{\hbar\pm}(z, u)$ in Proposition \ref{canonicalsol} satisfy the equation
$\nabla_{d}^{u_0} \mathcal{F}_{\hbar\pm}=0$.
\end{pro}
\pf Just use Lemma \ref{dependu} and the defining equation \eqref{gauge1} of $T(u)$. \qed

\vspace{1mm}

Note that the isomonodromy Casimir $\Omega(u)$, as well as the iKZ connection are also defined over neighbourhood of $u_0$ in $\h_{\rm reg}$. Although $\mathcal{F}_{\hbar\pm}(z, u)$ are defined for real $u$, we can take their analytic continuation to other points not necessary real. We shall call them the {\it canonical solutions} of the iKZ equations $\nabla^{u_0} \mathcal{F}_\hbar=0$ and $\nabla^{u_0}_{d} \mathcal{F}_\hbar=0$.

It also follows from Proposition \ref{solISO} that the ratio of the canonical solutions $\mathcal{F}_{\hbar\pm}$, i.e., the quantum Stokes matrices $S^{u_0}_{\hbar\pm}(u)$ of the iKZ connection, are locally constant.
Thus Sections \ref{sec:StokesMa} and \ref{sec:isodeformation} take respectively two equivalent approaches to the isomonodromy deformation problem in the spirit of Jimbo-Miwa-Ueno \cite[Section 3]{JMU}:
\begin{enumerate}
\item To start from a family of functions $\mathcal{F}_\hbar(z, u)$, parametrized by some $u$, having the monodromy/Stokes data, independent of $u$, and to derive a system of linear differential equations in $(z, u)$ for $\mathcal{F}_\hbar(z, u)$.

\item To construct non-linear differential equations on the space of singularity data, so that
each solution leaf (viewed as a family of ordinary differential equations) corresponds to one and the same
partial monodromy data.
\end{enumerate}

\subsection{Quantum Stokes matrices and Yang-Baxter equations}\label{sec:StokesR}
\begin{pro}For any $u\in\h_{\rm reg}(\mathbb{R})$, the quantum Stokes matrices $S_{\hbar\pm}(u)$ of the gKZ connection satisfy the Yang-Baxter equation 
\[S_\pm^{12}S_\pm^{13}S_\pm^{23}=S_\pm^{23}S_\pm^{13}S_\pm^{12}\in \UU^{\wh{\otimes} 3}.\]
Here we use the standard convention that $S^{12}:=\sum_a X_a\otimes Y_a\otimes 1$, $S^{13}:=\sum_a X_a\otimes 1\otimes Y_a$, $S^{23}:=\sum_a 1\otimes X_a\otimes Y_a$ for any $S=\sum_a X_a\otimes Y_a$.
\end{pro}
\pf 
It can be proved in the same way as the finite dimensional cases \cite{Xu3, Xu4}. \qed
\begin{rmk}
The same proposition first appears in \cite{TLXu} in a different setting (see Remark \ref{diffsolutions}), and appears in \cite{Xu3, Xu4} in a categorical setting.
\end{rmk}
As a consequence of the isomonodromy property, i.e., $S^{u_0}_{\hbar\pm}(u)=S_{\hbar\pm}(u_0)$ for any $u\in\mathcal{D}$, we have

\begin{thm}\label{YBeq}
The quantum Stokes matrices $S^{u_0}_{\hbar\pm}(u)$ of the iKZ connection satisfy the Yang-Baxter equation.
\end{thm}

\subsection{Generalized cyclotomic KZ connections and isomonodromy deformation}\label{sec:cycStokesMa}
This subsection concerns the Stokes phenomenon and isomonodromy deformation of the cyclotomic analog of gKZ connections. As we will see, they are parallel to the ones for gKZ connections. The results in this subsection are claimed in \cite{Xu4} (and are proved in the categorical setting).
\subsubsection{Admissibility and classical limit}
let us take the complex Lie algebra $\g={\rm gl}_n$, and take the negative transpose $\tau$ as an involution of $\g$ with the fixed point Lie algebra $\frak k={\rm so}_n$.

Set $\UU_\frak k=U(\frak k)\fml$ and $\UU=U(\frak g)\fml$. Let $\eps:\UU_\frak k\to\IC\fml$ be the counit of $\UU_\frak k$. Then $\UU_\frak k$ splits as
$\Ker(\eps)\oplus\IC\fml\cdot 1$, with projection onto the first summand denoted by $\pi$. 
Define $\UU_\frak k'\subset\UU_\frak k$ by
\[\label{qfsha}\UU_\frak k'=\{x\in\UU_\frak k|\, \pi^{\otimes n}\circ\Delta^{(n)}(x)\in\hbar^n\UU_\frak k^{\otimes n},\,n\geq 1\},\]
where recall $\Delta^{(n)}:\UU_\frak k\to\UU_\frak k^{\otimes n}$
is the iterated coproduct.
The algebra $\UU_\frak k'$ has a natural Hopf algebra structure, known as a quantum formal series Hopf algebra, and $\UU_\frak k'=U(\hbar\frak k\fml)$. Furthermore, $\UU'_\frak k$ is a flat deformation of the completed symmetric algebra $\wh
{S}\frak k=\prod_{n\geq 0}S^n\frak k$.

An element $A\in\UU_\frak k\wh{\otimes}\UU$ is called admissible, if $A$ is further inside the subalgebra $\UU_\frak k'\wh{\otimes}\UU$.
Given an admissible $A\in\UU'_\frak k\wh{\otimes}\UU$, the {\it classical limit}
of $A$, denoted by $\cl{A}$ is the image of $A$ in
\[\UU_\frak k'\wh{\otimes}\UU/\hbar(\UU'_\frak k\wh{\otimes}\UU)
=\wh{S}\frak k\wh{\otimes}\Ug\]
Given that $\wh{S}\frak k=\IC\llbracket \frak k^*\rrbracket $, we will regard $\cl{A}$ as formal
function on $\frak k^*$ with values in $\Ug$.

Similar to Section \ref{sec:scl}, we introduce the space $G\llbracket\frak k^*\rrbracket_0$ of group like elements in the classical limit, which embeds into the
topological Hopf algebra $\widehat{S}(\frak k)\wh{\otimes}\Ug=\lim_p(\IC [\frak k^*]/\mm^p)\otimes \Ug$. 

\subsubsection{Canonical solutions of generalized cyclotomic KZ equations}
Let $\{e_i\}_{i\in I}$ be an orthonormal basis of $\frak k={\rm so}_n$ with respect to the Killing form of $\g$, and let $\Omega_\frak k=\sum_{i\in I_+} e_i\otimes e_i\in \frak k\otimes \frak k$. Furthermore, let us denote the Casimir element by $C_\frak k=\sum_{i\in I}e_ie_i\in U(\frak k).$

The {\it generalized cyclotomic KZ (gcKZ) connection} with a parameter $u\in\frak h$ is the $\UU_\frak k\wh{\otimes}\UU$–valued connection on $\mathbb{C}$ given by
\begin{eqnarray}\label{cycdKZ}
\nabla\GCKZ=d_{z}-\Big(u^{(1)}+\frac{2\Omega_\frak k+C_\frak k^{(1)}}{z}\Big)\cdot dz,
\end{eqnarray}
where $u^{(1)}:=u\otimes 1$ and $C_\frak k^{(1)}:=C_\frak k\otimes 1$.
We will assume that $u\in\frak h_{\rm reg}(\mathbb{R})$, which determines the Stokes sectors ${\IH}_\pm$ as in Section \ref{sec:solgKZ}. Furthermore, similar to the gKZ equation in Section \ref{sec:solgKZ}, we can introduce the canonical solutions of $\nabla\GCKZ F_\hbar=0$ with prescribed asymptotics at $\infty$ in any $\IH_\pm$. The following theorem can be proved in a same way to Theorem \ref{th:Stokes}. We choose the determination of ${\rm log}z$ with a cut along $i\mathbb{R}_{<0}$. 
\begin{thm}\label{th:cycStokes}
For any $u\in\frak h_{\rm reg}(\IR)$, there are unique holomorphic functions
$H_{\hbar\pm}:\IH_\pm\to \UU_\frak k'\wh{\otimes}\UU$ such that $H_{\hbar\pm}(z,u)$ tends to $1$
as $z\to \infty$ within $\wh{\IH}_\pm$, and the $\UU_\frak k'\wh{\otimes}\UU$--valued
function
\[{F_\hbar}_\pm(z,u)=H_{\hbar\pm}(z,u)\cdot z^{\hbar C} e^{z u^{(1)}}\]
satisfies $\nabla\GKZ{F_\hbar}_\pm=0$. Here $C:=\sum_a e_ae_a\in U(\g)$ for any orthonormal basis $\{e_a\}$ of $\frak g$.
\end{thm} 
\begin{rmk}
For the appearance of the element $C$ in the expression, we refer to \cite{Xu4} (for a finite dimensional analog).
\end{rmk}
Let us take a root space
decomposition $\g = \mathfrak h \oplus_{\alpha\in\Phi}\mathbb{C}e_\alpha$. For any positive root $\alpha\in\Phi_+$, set $C_{\frak k,\alpha}=\frac{1}{2}(e_{\alpha}+\tau(e_{\alpha}))(e_{-\alpha}+\tau(e_{-\alpha}))$ (recall that $\tau$ is the involution on $\g$). An analog of Lemma \ref{dependu} is
\begin{lem}
The function ${F_\hbar}_\pm$ satisfies
\[\left(d_\h-\frac{\hbar}{2}\sum_{\alpha\in\Phi_+}\frac{d\alpha}{\alpha}
\Delta(C_{\frak k,\alpha})-z(d_\h u^{(1)})\right){F_\hbar}_\pm=
{F_\hbar}_\pm\left(d_\h-\frac{\hbar}{2}\sum_{\alpha\in\Phi_+}\frac{d\alpha}{\alpha}
(\onetwo{C_{\frak k,\alpha}})\right)\]
where $\Delta$ is the coproduct on $\UU$.
%
\end{lem}

\subsubsection{Quantum Stokes matrices}
Let $F_{\hbar\pm}$
be the canonical solutions of \eqref{cycdKZ} with respect to ${\rm Sect}_\pm$.

\begin{defi}
For any $u\in\h_{\rm reg}(\mathbb{R})$, the quantum Stokes matrices $K_{\hbar\pm}(u)\in\UU^{\otimes 2}$ of the gcKZ connection are defined by
\[{F_\hbar}_+={F_\hbar}_-\cdot e^{-\hbar\pi i C}K_{\hbar+}(u)\aand
{F_\hbar}_-={F_\hbar}_+\cdot K_{\hbar-}(u)e^{\hbar\pi i C}\]
where the first identity is understood to hold in $\IH_-$ after ${F_\hbar}
_+(z,u)$ has been continued across the ray $i\IR_{\geq 0}$, and the second
in $\IH_+$ after ${F_\hbar}_-(z,u)$ has been continued across $i\IR_{\leq 0}$. 
\end{defi}

\subsubsection{Isomonodromy cyclotomic KZ equations}
Let us consider the equation
\begin{eqnarray}\label{gauge2}
d_\frak h G(u)=\frac{\hbar}{2}\sum_{\alpha\in\sfPhi_+}\frac{d\alpha}{\alpha}(\onetwo{C_{\frak k,\alpha}})G(u).
\end{eqnarray}
Given any fixed initial point $u_0\in\h_{\rm reg}(\mathbb{R})$. Let $G_\frak k(u)\in \UU_\frak k\wh{\otimes}\UU$ be the solution of \eqref{gauge2} defined on a neighbourhood $\mathcal{D}$ of $u_0$ in $\h_{\rm reg}(\mathbb{R})$ with $G(u_0)=1\otimes 1$.

\begin{defi}
The isomonodromy cyclotomic KZ (icKZ) connection, with respect to the chosen point $u_0$, is the $\UU_\frak k\wh{\otimes}\UU $-valued flat connection on $\IC\times\mathcal{D}$ given by
\begin{eqnarray}
&&\nabla_{c}^{u_0}= d_z-\left(u^{(1)}+\hbar\frac{2\Omega_\frak k(u)+C_\frak k^{(1)}}{z}\right)dz,\label{cycisoKZ1}\\
&&\nabla_{cd}^{u_0}=d_\frak h-\left( z d_\h u^{(1)}+\hbar\sum_{\alpha\in\sfPhi_+}\frac{d\alpha}{\alpha}G^{-1}(u)\Omega_{\frak k,\alpha}G(u)\right).\label{cycisoKZ2}
\end{eqnarray}
Here the subscript $c$ stands for "cyclotomic", and $cd$ stands for "cyclomotic dynamical".
\end{defi}
The flatness of the connection comes from the defining equation \eqref{gauge2} of $G(u)$.
\begin{pro}\label{cyccanonicalsol}
For any fixed $u\in\mathcal{D}$, let $H_{\hbar\pm}$ be the holomorphic functions as in Theorem \ref{th:cycStokes}.
Then $\mathcal{H}_{\hbar\pm}(z,u):=G(u)^{-1}H_{\hbar\pm}(z,u)G(u)$ are the unique holomorphic functions on $\wh{\IH}_\pm$ with valued in $\UU'_\frak k\wh{\otimes}\UU$ such that $\mathcal{H}_{\hbar\pm}(z,u)$ tends to $1$
as $z\to \infty$ within $\wh{\rm Sect}_\pm$, and the
function
\[\mathcal{F}_{\hbar\pm}(z,u)=\mathcal{H}_{\hbar\pm}(z,u)\cdot z^{\hbar C} e^{z u^{(2)}}\]
satisfies the icKZ equations $\nabla_c^{u_0} \mathcal{F}_\hbar=0$ and $\nabla_{cd}^{u_0} \mathcal{F}_\hbar=0$.
\end{pro}
This proposition is similar to Proposition \ref{canonicalsol} and \ref{solISO}, so does the proof. We shall call $\mathcal{F}_{\hbar\pm}$ the {\it canonical solutions} of the icKZ equations.

\begin{thm}\label{thm:cycqisomono}
The quantum Stokes matrices $K^{u_0}_{\hbar\pm}(u)$ of the icKZ connection stay constant in $\mathcal{D}$ (independent of $u$).
\end{thm}
\pf The ratio of the two solutions $\mathcal{F}_{\hbar\pm}$ of the common linear differential equation $\nabla^{u_0}_{cd} \mathcal{F}_\hbar=0$ doesn't depend on $u$ in $\mathcal{D}$.
\qed 

\subsubsection{Quantum Stokes matrices and reflection equations}\label{sec:cycStokesR}
It follows from \cite{Xu4} that
\begin{thm}\label{reflectioneq}
For any $u\in\h_{\rm reg}(\mathbb{R})$, the quantum Stokes matrices $K_{\hbar+}(u)\in \UU_\frak k\wh{\otimes}\UU$ and
$S_{\hbar+}(u)\in \UU^{\wh{\otimes} 2}$ (of the gKZ and gcKZ connections) satisfies the $\tau$-twisted reflection equation
\begin{eqnarray}
K_{\hbar+}^{12}(\tau S_{{\hbar+}})^{32} K_{\hbar+}^{13}S_{\hbar+}^{32}=S_{\hbar+}^{32}K_{\hbar+}^{13}(\tau S_{{\hbar+}})^{23} K_{\hbar+}^{12}\in \UU_\frak k\wh{\otimes}\UU^{\wh{\otimes} 2}.
\end{eqnarray}
Here $\tau S_{\hbar+}:=(\tau\otimes 1)(S_{\hbar+})$, and the involution $\tau$ of $\g$ extends to an automorphism of $U(\g)$. 
\end{thm}
As an immediate consequence of the isomonodromy property, i.e., $K^{u_0}_{\hbar+}(u)=K_{\hbar+}(u_0)$, we have
\begin{thm}\label{RFeq}
The quantum Stokes matrix $K^{u_0}_{\hbar+}(u)$ of the icKZ connection satisfy the $\tau$-twisted reflection equation.
\end{thm}

\section{Classical limit}\label{sec:sclofKZ}
In this section, we show that the classical limit of the iKZ (resp. icKZ) connection gives the Dubrovin connections of semisimple Flat F-manifolds (resp. Frobenius manifolds). In Section \ref{sec:hadic}, we study the $\hbar$-adic property of isomonodromic KZ connections. In Section \ref{sec:CasimirIso}--\ref{qmonotomono}, we show that the classical limit of (the monodromy of) iKZ connections 
coincides with (the monodromy of) Dubrovin connections.

\subsection{$\hbar$-adic property of the iKZ connections}\label{sec:hadic}
Let us write the solution of the equation \eqref{gauge1}
\[d_\frak h T(u)=\frac{\hbar}{2}\sum_{\alpha\in\sfPhi_+}\frac{d\alpha}{\alpha}(\onetwo{C_\alpha})\cdot T(u),\]
as $T(u)=e^{E(u)}$, where $E(u)=\hbar E_1(u)+\hbar^2 E_2(u)+\cdot\cdot\cdot$ is the Magnus expansion \cite{Magus}. Then each $E_i$ is given by an iterated integral as follows.

Let us take $u_0\in\h_{\rm reg}(\mathbb{R})$ and $\mathcal{D}$ a neighbourhood of $u_0$ in $\h_{\rm reg}$. Let $I:[0,1]\rightarrow \mathcal{D}$ be a path from $u_0$ to any $u\in\mathcal{D}$. We denote by $A(t)dt$ (a 1-form valued in $U(\g)^{\otimes 2}$) the pull back of the 1-form $\sum_{\alpha\in\sfPhi_+}\frac{d\alpha}{\alpha}(\onetwo{C_\alpha})$ on $\mathcal{D}$ by $I$. It follows from the continuous Baker-Campbell-Hausdorff formula (also known as generalized Baker-Campbell-Hausdorff-Dynkin, see, e.g., \cite{Strichartz}) that 
\begin{eqnarray*}
&&E_1=\int_{0}^1 dt_1A(t_1),\\
&&E_2=\frac{1}{2}\int_{0}^1 dt_2\int_{0}^{t_2}dt_1[A(t_2),A(t_1)],\\
&&E_3=\frac{1}{6}\int_{0}^1dt_3\int_{0}^{t_3}dt_2\int_{0}^{t_2}dt_1\left([A(t_3),[A(t_2),A(t_1)]]+[A(t_1),[A(t_2),A(t_3)]]\right),\\
&&......\end{eqnarray*}
where the $i$-th order term $E_i$ is represented as an iterated integral of a linear combination of the nested commutators of $n$ $A(t_i)'s$. In particular, because the standard order filtration degree of $A(t)$ in $U(\g)^{\otimes 2}$ is 2, the filtration degree of $E_i$ in $U(\g)^{\otimes 2}$ is less than or equal to $i+1$. That is, if we write $E(u)=\sum_{i\geq 0}\hbar^i x_i$, then the filtration order of $x_i$ is less than or
equal to $i+1$.

Now we show the $\hbar$-adic property of the isomonodromic Casimir element $\Omega(u)$ (with respect to $u_0$).
\begin{pro}\label{pro:hadic}\hfill
\begin{enumerate}
    \item For any $X\in \UU'\wh{\otimes}\UU$ and $u\in\mathcal{D}$, we have $T(u)^{-1}X T(u)\in \UU'\wh{\otimes}\UU$. In particular, $\hbar\Omega(u)=T(u)^{-1}(\hbar\Omega) T(u)\in \UU'\wh{\otimes}\UU$;
    \item the classical limit $I(u)$ of $\hbar\Omega(u)$ is a (formal) function on $\g^*$ valued in $\g\subset U(\g)$. 
\end{enumerate}
\end{pro}
\pf 
1. Recall that $T(u)=e^{E(u)}$, where $E(u)$ is the Magnus expansion, and the filtration degree of $x_i\in U(\g)^{\otimes 2}$ in $E(u)=\sum_{i\geq 0}\hbar^i x_i$ is less than or equal to $i+1$. On the other hand, taking the Lie algebra $\UU^{\wh{\otimes}2}$ for the commutator, we have $$
T(u)^{-1}XT(u)= e^{-E(u)}Xe^{E(u)}=X+[E(u),X]+\cdot\cdot\cdot+\frac{1}{n!}[E(u),[E(u),...,[E(u),X]...]+....
$$
Therefore by the above identity and Lemma \ref{hbaradic}, $T(u)^{-1}XT(u)\in \UU'\wh{\otimes}\UU$.

2. From the discussion in part 1, we see that the nonzero contributions to the classical limit $I(u)$ of $\hbar\Omega(u)=e^{-E(u)}(\hbar\Omega) e^{E(u)}$ are from the terms $E^{(2)}(u)=\hbar E^{(2)}_1(u)+\hbar^2 E^{(2)}_2(u)+...$ in $E(u)$, where for example
\begin{eqnarray*}
&&E^{(2)}_1=\int_{0}^1 dt_1A^{(2)}(t_1),\\
&&E^{(2)}_2=\frac{1}{2}\int_{0}^1 dt_2\int_{0}^{t_2}dt_1[A^{(2)}(t_2),A^{(2)}(t_1)],\\
&&......\end{eqnarray*}
for $A(t)^{(2)}dt$ the pull back of the 1-form $\sum_{\alpha\in\sfPhi_+}\frac{d\alpha}{\alpha}C_\alpha^{(2)}$ on $\mathcal{D}$. In other words, $e^{E^{(2)}(u)}$ is the solution of the equation $d_\frak h T(u)=\hbar\sum_{\alpha\in\sfPhi_+}\frac{d\alpha}{\alpha}C_\alpha^{(2)}\cdot T(u)$ with initial condition $T(u_0)=1$.

Hence we deduce that
\[(\Delta\otimes 1)(I(u))=(\Delta\otimes 1)\scl{e^{-E^{(2)}(u)}(\hbar\Omega) e^{E^{(2)}(u)}}=\scl{e^{-E^{(2)}(u)}(\hbar(\Delta\otimes1)(\Omega)) e^{E^{(2)}(u)}}.\]
Here $\Delta$ is the coproduct.
It then follows from $(\Delta\otimes1)(\Omega)=\Omega^{13}+\Omega^{23}$ that $$(\Delta\otimes 1)(I(u))=I(u)^{13}+I(u)^{23},$$ which means that the image of $I(u)\in \widehat{S}(\g)\wh{\otimes}U(\g)$, though as formal function from $\g^*$ to $U(\g)$, is primitive, i.e., is valued in $\g\subset U(\g).$ 
\qed

\begin{ex}
By definition, $\hbar\Omega(u_0)=\hbar\Omega$, thus $I(u_0)={\rm cl}(\hbar\Omega(u_0))$ coincides with the isomorphism $\g^*\cong\g$ given by an invariant product on $\g$.
\end{ex}

\subsection{Classical limit of isomonodromy Casimir}\label{sec:CasimirIso}
By Proposition \ref{pro:hadic}, the classical limit $I(u)$ of $\hbar\Omega(u)$ is a map $I(u):\mathcal{D}\times\g^*\rightarrow\g$.

\begin{pro}\label{isoCasimir}
For any $V_0\in\g^*$, the function $I(u;V_0):\mathcal{D}\rightarrow \g$ is the solution of the equation \eqref{simplepole} with the initial condition $I(u_0;V_0)=V_0$ at $u_0\in\mathcal{D}$.
\end{pro}
\pf By Theorem \ref{thm:qisomono}, the quantum Stokes matrices $S^{u_0}_{\hbar\pm}(u)$ of
\[\nabla^{u_0}=d_z-\Big(u^{(2)}+\hbar\frac{\Omega(u)}{z}\Big)dz\]
are preserved. Here we have assumed the initial condition $\Omega(u_0)=\Omega$. Thus by taking classical limit and evaluating on $V_0\in \g\cong\g^*$, we conclude that the Stokes matrices $S_\pm(I(u;V_0))$ of \[\nabla_z=d_z-\left(u+\frac{I(u;V_0)}{z}\right)dz\] are preserved, i.e., don't depend on $u$.

On the other hand, following \cite{Boalch2}, equation \eqref{simplepole} describes exactly the isomonodromy deformation for $G$-valued Stokes phenomenon. Thus $I(u;V_0)$ is a solution of \eqref{simplepole} with the initial condition $I(u_0;V_0)=V_0$ at $u_0\in\mathcal{D}$.
\qed

\subsection{Classical limit of iKZ connections}\label{sec:sclDubrovin}
We have seen from the proof of Theorem \ref{th:Stokes} that
if $F_\hbar(z)\in \UU\otimes\UU'$ is a solution of the gKZ equation
\[\frac{d F_\hbar}{dz}=
\left(u^{(2)}+\hbar\frac{\Omega}{z}\right)F_\hbar,\]
then the classical limit $F(z)\in G\llbracket\frak g^*\rrbracket_0$
of $F_\hbar$ satisfies
\[\frac{dF(z;V)}{dz}=
\left(u+\frac{V}{z}\right)F(z;V),\]
for any $V\in \g^*\cong\g$. Given Proposition \ref{isoCasimir}, the similar result for iKZ equation can be described as follows.

For any $V_0\in\g$, we denote by $\nabla_{V_0}$ the flat connection of flat F-manifolds parameterized by $V_0$ (see Section \ref{Frobeniusmanifold})
\begin{eqnarray*}
&&\nabla_z=d_z -\left(u+\frac{V(u)}{z}\right)dz,\\
&&\nabla_u=d_\frak h -\left( z d_\frak hu+\Lambda(u)\right),
\end{eqnarray*}
where $V(u)$ is the solution of \eqref{simplepole} with the initial value $V(u_0)=V_0$. 

\begin{pro}\label{mainthm}
For any $u\in \h_{\rm reg}(\IR)$, let $\mathcal{F}_{\hbar\pm}(z,u)$ be the canonical solutions of $\nabla^{u_0} F_\hbar=0$ given in Proposition \ref{canonicalsol}. Then the classical limit $F_\pm:=\scl{F_{\hbar\pm}}$ take values in $G\llbracket\g^*\rrbracket_0$, and for any $V_0$, $F_\pm(z,u;V_0)$ coincide with the canonical solutions (i.e., solutions with the prescribed asymptotics at irregular singularity) of the equations $\nabla_{V_0} F=0$.
\end{pro}
\subsection{Classical limit of quantum Stokes matrices}\label{qmonotomono}
The following result is a direct consequence of Proposition \ref{mainthm}. 
\begin{pro}\label{qStoS}
The classical limit ${\rm cl}(S^{u_0}_{\hbar\pm})$ of the quantum Stokes matrices $S^{u_0}_{\hbar\pm}\in \UU'\wh{\otimes}\UU$ takes values in
$G\llbracket\g^*\rrbracket_0$, and for any $V_0$, ${\rm cl}(S^{u_0}_{\hbar\pm})(u;V_0)\in G$ coincides with the Stokes matrices of the
connection $\nabla_{V_0}$.
\end{pro}

In summary, the classical limit of the canonical solutions and Stokes matrices of the iKZ connections recover the counterparts of the flat connections of flat F-manifolds, associated to a real semisimple point $u_0\in\h_{\rm reg}(\mathbb{R})$. Furthermore, these connections have the same isomonodromy property (Theorem \ref{thm:qisomono} and Proposition \ref{sclisomono}).

\subsection{Classical limit of icKZ connections}\label{sec:clicKZ}
This subsection is an exposition of the results in \cite{Xu4}. It shows that the classical limit of the icKZ connection coincides with the Dubrovin connections of semisimple Frobenius manifolds. The proofs of all the results are similar to the counterparts of iKZ connections. 

First analogue to Section \ref{sec:hadic}, one can show that the residue $2G_\frak k(u)+C_\frak k^{(1)}$ of the connection $\nabla^{u_0}_c$ (with respect to $u_0$) in \eqref{cycisoKZ1} is inside the subspace $\UU'_\frak k\wh{\otimes}\UU$, and its classical limit $I_\frak k(u)\in \wh{S}(\frak k)\wh{\otimes}U(\g)$ is actually a map $I(u):\mathcal{D}\times\frak k^*\rightarrow\g\subset U(\g)$.

\begin{pro}\label{isoCasimir1}
For any $V_0\in\frak k^*$, the function $I_\frak k(u;V_0):\mathcal{D}\rightarrow \g$ is the solution of the isomonodromy equation \eqref{simplepole} with the initial value $I(u_0;V_0)=V_0\in \frak k$.
\end{pro}
It follows from this proposition that the classical limit of the icKZ connections coincides with the flat connections $\nabla_{V_0}$ for all $V_0\in\frak k$. Furthermore one can show that
\begin{pro}\label{thm:cycscl}
For any fixed $u\in \h_{\rm reg}(\IR)$, let $\mathcal{F}_{\hbar\pm}(z,u)$ be the canonical solutions of $\nabla_c^{u_0} F_\hbar=0$ given in Proposition \ref{cyccanonicalsol}. Then the classical limit $F_\pm:=\scl{F_{\hbar\pm}}$ take values in $G\llbracket\g^*\rrbracket_0$, and for any $V_0\in\frak k$, $F_\pm(z,u;V_0)$ coincide with the canonical solutions (i.e., solutions with the prescribed asymptotics at irregular singularity) of the equations $\nabla_{V_0} F=0$.
\end{pro}

\begin{pro}\label{clKmatrix}
The classical limit ${\rm cl}(K^{u_0}_{\hbar\pm})$ of the quantum Stokes matrices $K^{u_0}_{\hbar\pm}\in \UU'_\frak k\wh{\otimes}\UU$ takes values in
$G\llbracket\g^*\rrbracket_0$, and for any $V_0$, ${\rm cl}(K^{u_0}_{\hbar\pm})(u;V_0)\in G$ coincides with the Stokes matrices of the connection $\nabla_{V_0}$.
\end{pro}

Recall from Section \ref{iviso} that for any $V_0\in\frak k$, $\nabla_{V_0}$ is the Dubrovin connection of the germ of semisimple Frobenius manifold at $u_0$ parameteried by $V_0$. In summary, the icKZ connection can be seen as a quantization of the Dubrovin connections. 

\begin{rmk}
We have seen that the classical limit of the icKZ connection coincides with the Dubrovin connections. In particular, Proposition \ref{isoCasimir1} says that the renormalization \[\Omega_\frak k(u)^\circ=V(u)+\hbar V_1(u)+o(\hbar)\in (\UU'_\frak k\wh{\otimes}\UU)\llbracket\hbar\rrbracket,\]
where $V(u)$ is the residue in Dubrovin connections. Recall that the Frobenius manifold structures near $u_0$ can be reconstructed by $V(u)$. Geometrically $\Omega_\frak k(u)$ would produce an $\hbar$-deformation of the datum of Frobenius manifolds (which is not in the category of Frobenius manifolds any more), and it is interesting to study this deformation in details.
\end{rmk}

\subsection{Quantization as a lifting problem}\label{qFM}
In this subsection, we explain the quantization of semisimple Frobenius manifolds as a lifting problem. The cases of flat F-manifolds will be the same.

Let us study the semisimple germs at the chosen point $u_0\in\h_{\rm reg}(\mathbb{R})$, once fixed the only variable is the initial value $V_0=V(u_0)$ of the isomonodromy equation \eqref{V(u)property}. As in Section \ref{iviso}, we denote by $\nabla_{V_0}$ the corresponding Dubrovin connection. As in Section \ref{sec:sol}, let us take any canonical solution $F_{V_0}(z,u)$ of $\nabla_{V_0}F=0$ on $\IC\times\mathcal{D}$, where $\mathcal{D}$ is a neighbourhood of $u_0$ in $\h_{\rm reg}(\mathbb{R})$. Then the monodromy property of $F_{V_0}$ (on the universal covering of $\mathbb{P}^1\setminus \{0,\infty\}$) in turn determines the connection $\nabla_{V_0}$, see e.g. \cite[Proposition 2.5 and 2.6]{JMU} and \cite[Lecture 4]{Dubrovin2}. Hence, the semisimple germs at $u_0$ is equivalently described by the collection of the canonical solutions $F_{V_0}$ for all $V_0$.
We thus obtain a map encoding the semisimple germs \[\tilde{F}(z,u):\frak k\cong\frak k^*\rightarrow {\rm GL}_n; \ V_0\mapsto F_{V_0}(z,u).\] 
Here recall that $\frak k={\rm so}_n$ and is identified with $\frak k^*$. Taking its Taylor expansion around $V_0=0\in\frak k$, we get a function valued in the formal Taylor series group ${\rm GL}_n\llbracket \frak k^*\rrbracket_0$, i.e., 
\[\tilde{F}:\IC\times\mathcal{D}\rightarrow {\rm GL}_n\llbracket\frak k^*\rrbracket_0\hookrightarrow \wh{S}(\frak k)\wh{\otimes}U(\g).\]

Since the classical limit of $\UU'_\frak k\wh{\otimes}\UU$ is $\wh{S}(\frak k)\wh{\otimes}U(\g)$, it motivates 
\begin{defi}\label{def:quantization}
A quantization of $\tilde{F}$ is a holomorphic function $F_{\hbar}$ on $\IC\times\mathcal{D}$ with values in $\UU'_\frak k\wh{\otimes}\UU$ such that $\scl{F_{\hbar}}=\tilde{F}$, i.e., $F_\hbar$ is a lift
\begin{equation*}
		\vcenter{\xymatrix{
		& &\UU'_\frak k\wh{\otimes}\UU \ar[d]^{c.l}\\
		\IC\times\mathcal{D} \ar@{-->}[rru]^{F_{\hbar}} \ar[rr]^{\tilde{F}} &&\UU'_\frak k\wh{\otimes}\UU/\hbar \UU_\frak k\wh{\otimes}\UU'
		}},
	\end{equation*}
and $F_\hbar(z,u)$ has the same monodromy property as $\tilde{F}$. 
\end{defi}
Here by the monodromy property, we refer to \cite[Proposition 2.5]{JMU}. It requires that, for example, there exists a constant element $S_\hbar\in \UU'_\frak k\wh{\otimes}\UU$, such that for any $u$, the functions $F_\hbar(z,u)$ and $F_\hbar(z,u) \cdot S_\hbar$ on the $z$-plane have the same aympototic expansion in the Stokes sector (defining $\tilde{F}$) and its opposite Stokes sector respectively. Taking classical limit, it implies that $\tilde{F}$ and $\tilde{F} \scl{S_\hbar}$ have the same aympototic expansion in the two opposite Stokes sectors. It then follows from definition that for any $V_0\in\frak k^*\cong\frak k$, $\scl{S_\hbar}(V_0)$ is one of the Stokes matrices of the Dubrovin connection $\nabla_{V_0}$. Therefore, the quantization of $\tilde{F}$ also encodes a deformation of the Stokes/monodromy data of the Frobenius manifolds.

In particular, the results in Section \ref{sec:clicKZ} imply
\begin{pro}\label{maincor}
The canonical solutions $\mathcal{F}_{\hbar\pm}$ of the icKZ equation, given in Proposition
\ref{cyccanonicalsol}, are the quantization of the functions $\tilde{F}_\pm:\IC\times\mathcal{D}\rightarrow {\rm GL}_n\llbracket\frak k^*\rrbracket_0$.
\end{pro}

Note that any element $A\in \widehat{S}(\frak k)\widehat{\otimes}U(\g)$, viewed as a formal function on $\frak k^*$, has a natural lift 
\[\hat{A}\in  \widehat{S}(\hbar\frak k)\widehat{\otimes}U(\g)\subset (\widehat{S}(\hbar\frak k)\widehat{\otimes}U(\g))\llbracket\hbar\rrbracket\cong \UU'_\frak k\wh{\otimes}\UU\] given by $\hat{A}(x):=A(\hbar x), \ \forall \ x\in\frak k^*.$ Here the isomorphism is through the PBW map. Since the product in $\UU'_\frak k$ is not commutative anymore, the natural lift of $\tilde{F}$ violates the second condition in Definition \ref{def:quantization}. Thus a quantization is a nontrivial correction of the natural lift. From the proof of Theorem \ref{th:Stokes}, we see that the desired correction terms in the power of $\hbar$ can be obtained by solving certain (inhomogeneous) ordinary differential equations in a recursive way.

\subsection{Quantum algebras and Poisson structures on moduli spaces}
In this subsection, we incorporate the Poisson structures on the moduli spaces of semisimple Frobenius manifolds and flat F-manifolds into the framework of the quantization.

It has been explained in our work \cite{Xu4} that the quantum Stokes data of the icKZ connection gives rise to a transcendental construction of a quantum symmetric pair in type AI, where the quantum Stokes matrix $K^{u_0}_\hbar$ plays the role of a universal K-matrix (for any $u_0\in\h_{\rm reg}(\mathbb{R})$). On the one hand, following Ciccoli and Gavarini \cite{CGa}, one can show that the classical limit of the quantum symmetric pair is isomorphic to the Dubrovin-Ugaglia Poisson space. On the other hand, by Proposition \ref{clKmatrix}, the classical limit of the K-matrix $K^{u_0}_\hbar$ coincides with the Riemann-Hilbert-Birkhoff map $\nu(u_0)$ in Theorem \ref{DUPoisson}. Furthermore, it explains the Poisson geometric nature of the map $\nu(u_0)$ from the quantum algebra aspect. See \cite{Xu4} for more details. Thus we have the following diagram

\begin{equation*}
    \begin{tikzcd}[column sep=4em, row sep=4em]
      \framebox{\Longstack[c]{Isomonodromy \\ cyclotomic KZ connections} } \arrow[r, "\mbox{ Stokes data}","\mbox{Thm \ref{RFeq}}"'] \arrow[d,shift left,"\mbox{\Longstack[c]{Classical limit\\ Pro \ref{thm:cycscl}}}"]  & \framebox{\Longstack[c]{Quantum symmetric pairs\\ K-matrices}} \arrow[d,shift left,"\mbox{\Longstack[c]{ Classical limit \\ Pro \ref{clKmatrix}}}"] \\ \framebox{\Longstack[c]{Dubrovin connections\\ Semisimple Frobenius manifolds}}  \ar[u,shift left,"\mbox{\Longstack[c]{Quantization}}"] \arrow[r, "\mbox{Stokes data}","\mbox{Thm \ref{DUPoisson}}"'] & \framebox{\Longstack[c]{Dubrovin-Ugaglia Poisson space \\ Riemann-Hilbert-Birkhoff maps}}
    \end{tikzcd}
\end{equation*}
\vspace{2mm}
Similarly, for semisimple flat F-manifolds, we have the diagram
\begin{equation*}
    \begin{tikzcd}[column sep=4em, row sep=4em]
      \framebox{\Longstack[c]{Isomonodromy \\ KZ connections} } \arrow[r, "\mbox{Stokes data}","\mbox{Thm \ref{YBeq}}"'] \arrow[d,shift left,"\mbox{\Longstack[c]{ Pro \ref{mainthm}}}"]  & \framebox{\Longstack[c]{Quantum groups\\ R-matrices}} \arrow[d,shift left,"\mbox{\Longstack[c]{Pro \ref{qStoS}}}"] \\ \framebox{\Longstack[c]{Flat connections\\ Flat F-manifolds}}  \ar[u,shift left,"\mbox{\Longstack[c]{Quantization}}"] \arrow[r, "\mbox{Stokes data}","\mbox{Thm \ref{BPoisson}}"'] & \framebox{\Longstack[c]{Dual Poisson groups\\Dual exponential maps}}
    \end{tikzcd}
\end{equation*}
Here we remark that (the boxes and arrows in) the first diagram can be seen as the "fixing locus" of the second diagram under the involution $\tau$ in various settings. We also remark that the theory of classical dynamical r-matrices and vertex-IRF transformation are closely related to the discussion in this subsection, following the work of \cite{EE}\cite{EEM}\cite{Xu1}. See the discussion in \cite{Xu4}.

\Addresses
\end{document}